\documentclass[lettersize,journal]{IEEEtran}
\usepackage{amsmath,amsfonts,amsthm}

\usepackage{algorithmic}
\usepackage{algorithm}
\usepackage{array}
\usepackage{subfigure}
\usepackage{textcomp}
\usepackage{stfloats}
\usepackage{url}
\usepackage{verbatim}
\usepackage{graphicx}
\graphicspath{{image/}}
\usepackage{cite}
\usepackage{booktabs}
\begin{document}

\title{MambaStock: Selective state space model for stock prediction}

\author{Zhuangwei Shi
\thanks{Correspondence author: Zhuangwei Shi (email:zwshi@mail.nankai.edu.cn)}
\thanks{Zhuangwei Shi is with the Ji Hua Laboratory, Foshan, Guangdong, China.}}

\markboth{Journal of \LaTeX\ Class Files,~Vol.~14, No.~8, August~2021}%
{Shell \MakeLowercase{\textit{et al.}}: A Sample Article Using IEEEtran.cls for IEEE Journals}


\maketitle

\begin{abstract}
    The stock market plays a pivotal role in economic development, yet its intricate volatility poses challenges for investors. Consequently, research and accurate predictions of stock price movements are crucial for mitigating risks. Traditional time series models fall short in capturing nonlinearity, leading to unsatisfactory stock predictions. This limitation has spurred the widespread adoption of neural networks for stock prediction, owing to their robust nonlinear generalization capabilities. Recently, Mamba, a structured state space sequence model with a selection mechanism and scan module (S6), has emerged as a powerful tool in sequence modeling tasks. Leveraging this framework, this paper proposes a novel Mamba-based model for stock price prediction, named MambaStock. The proposed MambaStock model effectively mines historical stock market data to predict future stock prices without handcrafted features or extensive preprocessing procedures. Empirical studies on several stocks indicate that the MambaStock model outperforms previous methods, delivering highly accurate predictions. This enhanced accuracy can assist investors and institutions in making informed decisions, aiming to maximize returns while minimizing risks. This work underscores the value of Mamba in time-series forecasting. Source code is available at \url{https://github.com/zshicode/MambaStock}.
\end{abstract}


\section{Introduction}
\label{sec:introduction}

\IEEEPARstart{S}{tock} market plays an important role in the economic development. Due to the high return characteristics of stocks, the stock market has attracted more and more attention from institutions and investors. However, due to the complex volatility of the stock market, sometimes it will bring huge loss to institutions or investors. Considering the risk of the stock market, the research and prediction on the change of the stock price can avoid the risk for the investors.

The traditional time series model ARIMA (Autoregressive Integrated Moving Average Model) can not describe the nonlinear time series, and needs to satisfy many preconditions before modeling, and can not achieve remarkable results in the stock forecasting. In recent years, with the rapid development of artificial intelligence theory and technology, more and more researchers apply artificial intelligence method to the financial market. On the other hand, the sequence modeling problem, focusing on natural language sequences, protein sequences, stock price sequences, and so on, is important in the field of artificial intelligence research \cite{shi2021vgaelda,jin2021lpigac}. The most representative artificial intelligence method is neural networks, which are with strong nonlinear generalization ability.

Recurrent Neural Network (RNN) was adopted for analyzing sequential data via neural network architecture, and Long Short-Term Memory (LSTM) model is the most commonly used RNN. LSTM introduced gate mechanism in RNN, which can be seen as simulation for human memory, that human can remember useful information and forget useless information \cite{jin2021chinese,jin2022tlcrys}. Attention Mechanism \cite{transformer2017} can be seen as simulation for human attention, that human can pay attention to useful information and ignore useless information. Attention-based Convolutional Neural Networks (ACNN) are widely used for sequence modeling \cite{JIN2021265,lin2022dattprot} and complex dependency captureing \cite{jin2022nimgsa,shi2022graph}. Combining Attention-based Convolutional Neural Networks and Long Short-Term Memory, is a self-attention based sequence-to-sequence (seq2seq) \cite{seq2seq2014} model to encode and decode sequential data. This model can solve long-term dependency problem in LSTM, hence, it can better model long sequences. LSTM can capture particular long-distance correspondence that fits the sturcture of LSTM itself, while ACNN can capture both local and global correspondence. Therefore, this architecture is more flexible and robust.

Transformer \cite{transformer2017} is the most successful sequential learning self-attention based model. Experiments on natural language processing demonstrates that Transformer can better model long sequences. Bidirectional Encoder Representation Transformer (BERT) with pretraining \cite{Devlin2018BERT} can perform better than the basic Transformer. Pretraining  is a method to significantly improve the performance of Transformer (BERT).

The models mentioned above have been applied for stock prediction. Box and Jenkins \cite{boxjenkins1970} adopted ARIMA for time-series forecasting. ARIMA-NN \cite{ZHANG2003159} improved time-series forecasting using a hybrid ARIMA and neural network model. Shi et al. \cite{shi2022attclx} proposed an attention-based CNN-LSTM and XGBoost hybrid model for stock prediction. Recently, neural network-enhanced state space Kalman Filter models have been applied for time-series forecasting. TL-KF \cite{shi2021tlkf} proposed Kalman Filter along with LSTM and Transformer for stock prediction \cite{shi2022attclx}.

The Mamba model \cite{gu2023mamba} represents a significant advancement in the field of sequence modeling. This model outperforms the traditional approach by incorporating a structured state space sequence model (S4) with a selection mechanism and scan module, known as S6. The Mamba model excels at capturing nonlinear patterns in sequential data, which has historically been a challenge for traditional time series models. The core strength of Mamba lies in its ability to efficiently model sequences using a linear-time complexity, making it suitable for processing large-scale datasets. The innovative selection mechanism allows it to dynamically adapt to different patterns and structures within the data, enabling more accurate predictions. Additionally, the scan module enhances the capability by scanning through the state spaces to identify relevant information for making predictions. 

Due to its versatility and adaptability, Mamba have been a popular choice for various sequence modeling tasks. However, the application of Mamba in financial time series remains to be explored. Therefore, this paper proposes a Mamba-based model to predict the stock price, named MambaStock. The MambaStock model, a novel approach to stock price prediction, effectively mines historical stock market data to accurately forecast future prices, eliminating the need for meticulous feature engineering or extensive preprocessing. Empirical studies across various stocks demonstrate the superiority of MambaStock over traditional methods, delivering precise predictions that can significantly inform investment decisions. This precision is invaluable to investors and institutions seeking to maximize returns while minimizing risks, highlighting the immense potential of Mamba in time-series forecasting. The source code of this paper is available at \url{https://github.com/zshicode/MambaStock}. The data is downloaded from Tushare(\url{www.tushare.pro}). The stock price data on Tushare is with public availability.

\section{Materials and Methods}

\subsection{Structured state space sequence model (S4)}

State space model is inspired by solving ODE \cite{kalman1960,shi2022diff}. Structured State Space sequence model (S4) is a recently proposed sequence modeling architecture that leverages the power of state spaces and structured matrices for efficient and effective processing of sequential data. It combines principles from control theory, signal processing, and deep learning to address the challenges associated with traditional sequence modeling approaches.

In S4 model, the key idea is to represent the underlying dynamics of a sequence using a state space representation. This representation captures the evolution of the systematic state over time, allowing for efficient computation and storage. The state space is parameterized using structured matrices, which impose certain constraints on the parameters to enable efficient training and inference.

Let $X\in\mathbb{R}^{B\times L\times D}$, where $B,L,D$ denote batch size, time steps and dimension, respectively. For each batch and each dimension, let $x_t,h_t,y_t$ denote the input, hidden state and output at time $t=1,2,...,L$, respectively, S4 model can be written as
\begin{equation}
    h_t=Ah_{t-1}+Bx_t
\end{equation}
\begin{equation}
    y_t=Ch_t
\end{equation}

By discretization, let $\Delta$ denotes the sample time, then
\begin{equation}
    \bar A=\exp(\Delta A)
\end{equation}
\begin{equation}
    \bar B=(\Delta A)^{-1}(\exp(\Delta A)-I)\cdot\Delta B
\end{equation}
\begin{equation}
    h_t=\bar Ah_{t-1}+\bar Bx_t
\end{equation}

Here, $A\in\mathbb{R}^{N\times N},B\in\mathbb{R}^{N\times 1},C\in\mathbb{R}^{1\times N}$, $h_t$ is $N$-dimension vector, $x_t,y_t,\Delta$ are numbers. Let $A$ be diagonal, $A$ can be also be storaged in $N$-dimension vector. Thus, considering all dimensions, the data structure is $A\in\mathbb{R}^{D\times N},B\in\mathbb{R}^{D\times N},C\in\mathbb{R}^{D\times N},\Delta\in\mathbb{R}^{D}$.

\subsection{Mamba}

Recently, Mamba, a structured state space sequence model with a selection mechanism and scan module (S6), has emerged as a powerful tool in sequence modeling tasks. In S4, $A,B,C,\Delta$ is time-invariant at any time $t=1,2,...,L$. However, Mamba introduces a novel selection mechanism that allows the model to dynamically choose which parts of the input sequence are relevant for making predictions. This mechanism helps the model focus on important information while ignoring irrelevant or noisy data, leading to improved generalization and performance. In selection mechanism, $B\in\mathbb{R}^{B\times L\times N},C\in\mathbb{R}^{B\times L\times N},\Delta\in\mathbb{R}^{B\times L\times D}$ can be learned from $X\in\mathbb{R}^{B\times L\times D}$ using fully-connected layers. By discretization, for each batch and each dimension, let $x_t,h_t,y_t$ denote the input, hidden state and output at time $t=1,2,...,L$, respectively, S4 model can be written as
\begin{equation}
    h_t=\bar A_t h_{t-1}+\bar B_t x_t
\end{equation}
\begin{equation}
    y_t=C_t h_t
\end{equation}

The scan module operates by applying a set of learnable parameters or operations to each window of the input sequence. These parameters are typically learned during training and can include convolutions, recurrent connections, or other types of transformations. By sliding this window over the entire sequence, the scan module is able to capture patterns and dependencies that span multiple time steps.

\subsection{MambaStock}

The MambaStock model leverages the Mamba framework to predict future stock price movement rates based on historical market data. It employs a range of features including opening price, high price, low price, trading volume, trading value, turnover rate, volume ratio, price-earnings ratio (PE), price-book ratio (PB), price-sales ratio (PS), total shares, float shares, free float shares, total market value, and circulating market value across different trading dates.

The model begins by processing the historical data, which is fed into the Mamba model with $N=16$, to capture temporal dependencies and extract relevant information. The Mamba model, through its internal mechanisms, is able to effectively mine the patterns and relationships within the input data.

The output of the Mamba model is then reduced to a one-dimensional representation, reflecting the predicted stock price movement rate for each future date. Since the movement rates are expected to fall within the (-1, 1) interval, a hyperbolic tangent (tanh) activation function is applied to ensure the output remains within this range.

\begin{figure*}
    \centering
    \includegraphics[width=0.9\textwidth]{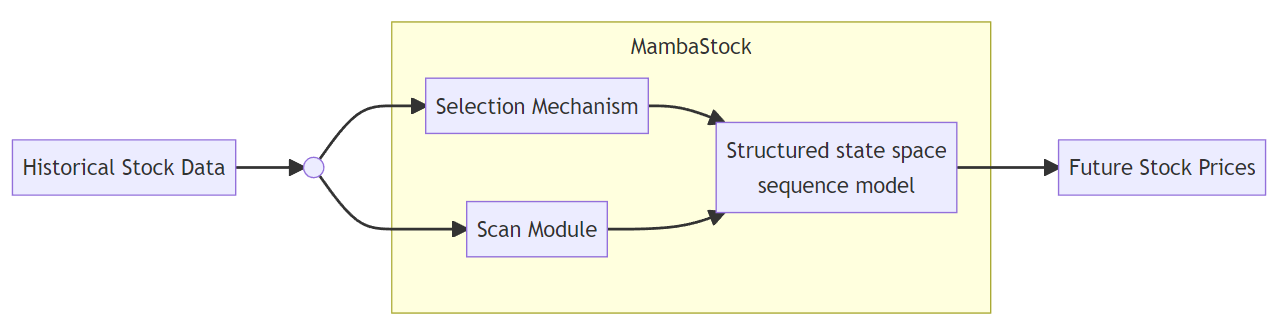}
    \caption{MambaStock framework.}\label{fig:structure}
\end{figure*}

To train the model, the mean squared error (MSE) is chosen as the loss function, as it measures the average squared difference between the predicted and actual stock price movement rates. By minimizing this loss, the model aims to improve its accuracy in predicting future stock price movements. The framework of MambaStock is shown on Fig. \ref{fig:structure}. 

The experiments are on an NVIDIA GTX3060 GPU with 12GB memory. The model is trained through Adam optimizer \cite{2014Adam}. The epoch number is 100 and learning rate is 0.01.

The data used in this article comes from the open and free public dataset in Tushare (\url{https://www.tushare.pro/}) for the research of stock market in China, which has the characteristics of rich data, simple use, and convenient implementation. It is very convenient to obtain the basic market data of stocks by calling the API.

\section{Experiments}

\subsection{Prediction performance}

\begin{figure}
    \centering
    \includegraphics[width=0.46\textwidth]{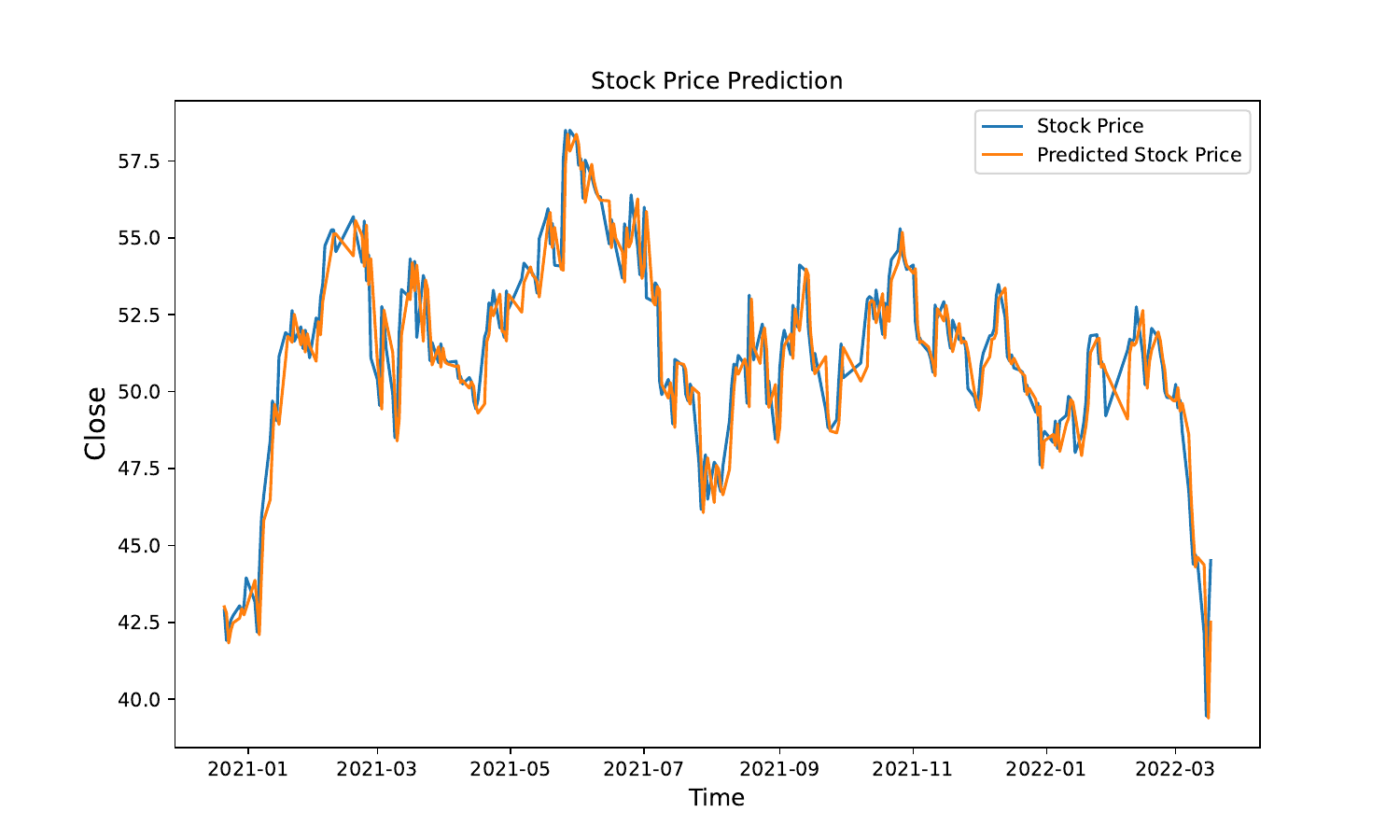}
    \caption{Prediction on 600036.SH}\label{fig:stock1}
\end{figure}

\begin{figure}
    \centering
    \includegraphics[width=0.46\textwidth]{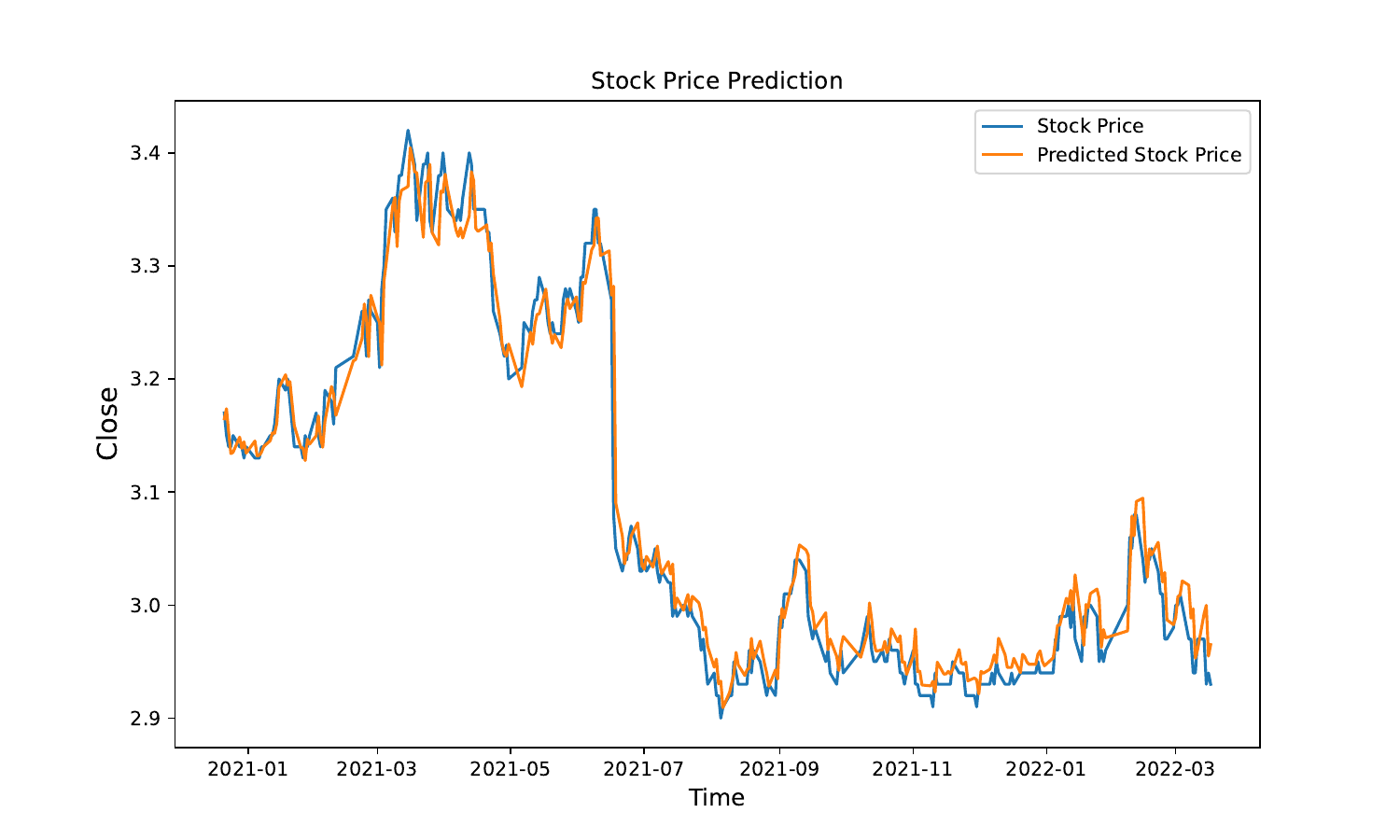}
    \caption{Prediction on 601288.SH}\label{fig:stock2}
\end{figure}

\begin{figure}
    \centering
    \includegraphics[width=0.46\textwidth]{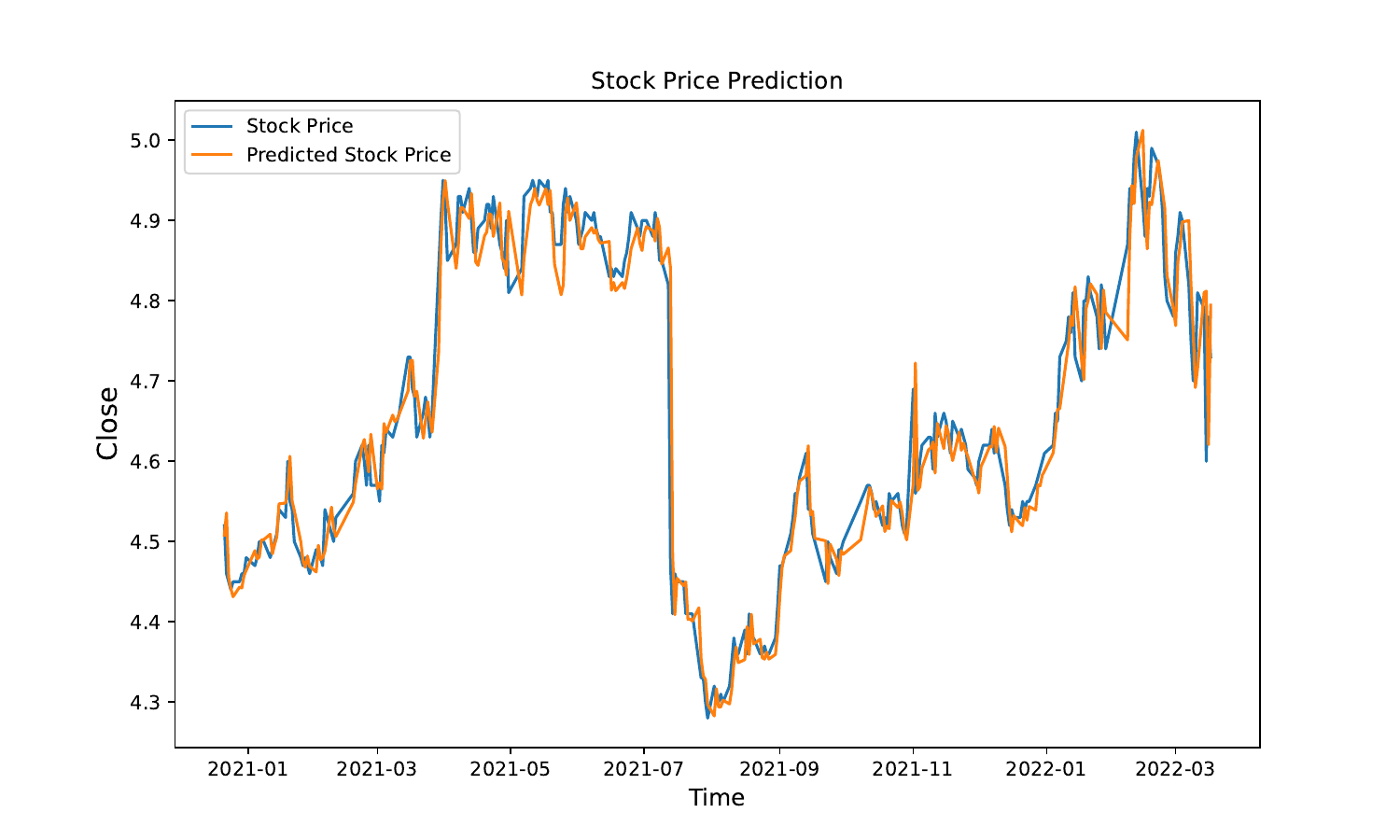}
    \caption{Prediction on 601328.SH}\label{fig:stock3}
\end{figure}

\begin{figure}
    \centering
    \includegraphics[width=0.46\textwidth]{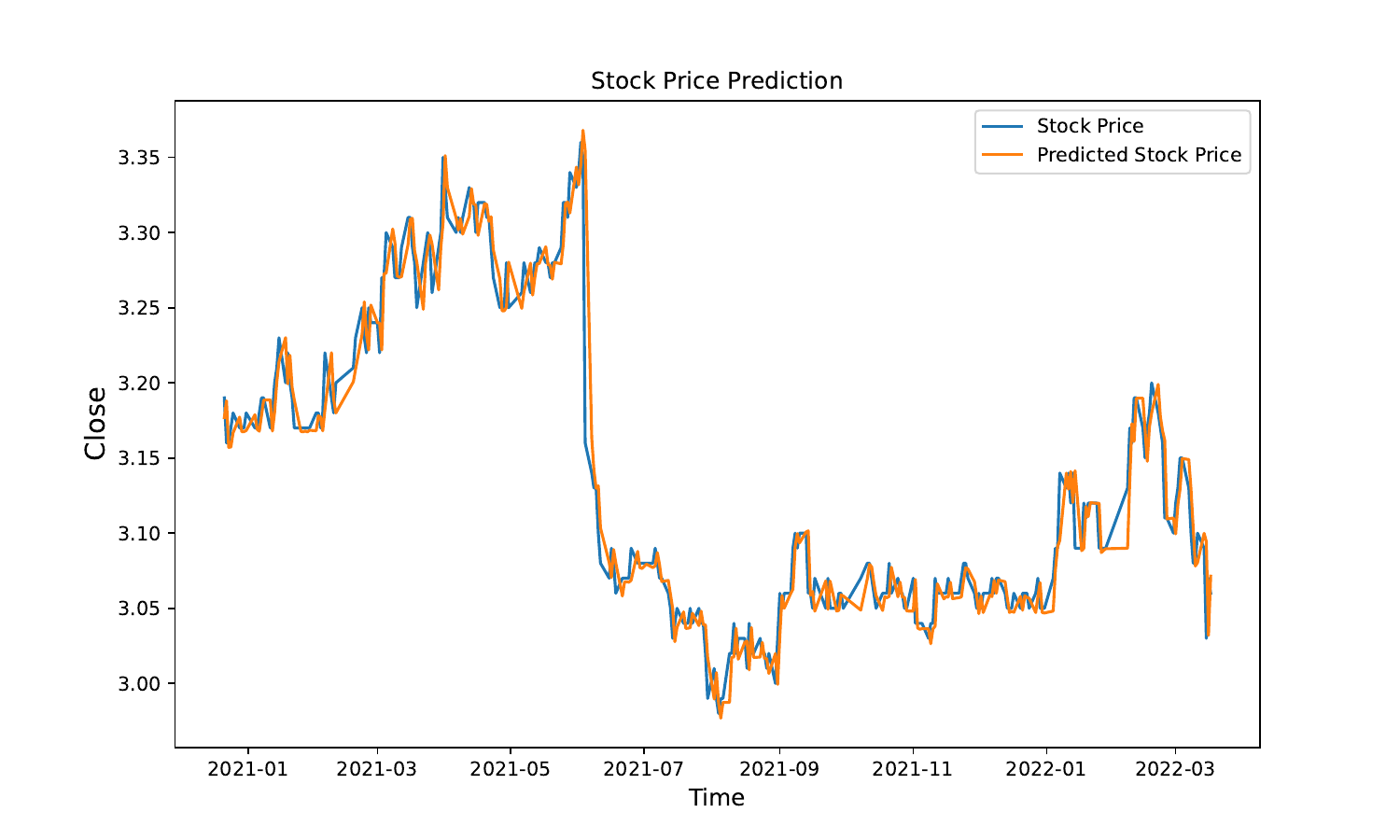}
    \caption{Prediction on 601988.SH}\label{fig:stock4}
\end{figure}

This paper conduct empirical study on the stock price of China Merchants Bank (600036.SH), the Agricultural Bank of China (601288.SH), Bank of Communications (601328.SH), and Back of China (601988.SH) in Chinese stock market. The data is downloaded from Tushare(\url{www.tushare.pro}). The stock price data on Tushare is with public availability. Regardless of the training set size, the test set size is limited to 300. The stock price prediction result of ARIMA model is shown on Fig. \ref{fig:stock1}-Fig. \ref{fig:stock4}. The results of MambaStock demonstrate its ability to accurately predict future stock prices. By leveraging historical market data and a range of financial indicators, the model has achieved significant accuracy in its predictions. The utilization of the Mamba framework allowed the model to capture temporal dependencies and extract relevant information effectively.

\subsection{Compared with other methods}

Then, empirical studies conduct comparison with current methods. The compared methods include:
\begin{itemize}
    \item KF: Kalman Filter, the traditional state space model.
    \item ARIMA: Box and Jenkins \cite{boxjenkins1970} adopted ARIMA for time-series forecasting.
    \item ARIMA-NN \cite{ZHANG2003159} improved time-series forecasting using a hybrid ARIMA and neural network model.
    \item XGBoost, single-directional LSTM, bi-directional LSTM (BiLSTM), and Transformer are also adopted as baselines.
    \item TL-KF \cite{shi2021tlkf} proposed Kalman Filter, along with LSTM and Transformer for stock prediction.
    \item AttCLX \cite{shi2022attclx} proposed an attention-based CNN-LSTM and XGBoost hybrid model for stock prediction.
\end{itemize}

The evaluation metrics are mean absolute error (MAE), root of mean square error (RMSE), mean absolute percentage error (MAPE) and $R^2$.
\begin{equation}
    MSE = \frac{1}{n}\sum\limits_{t = 1}^n {\left| {{{\hat X}_t} - {X_t}} \right|^2} 
\end{equation}
\begin{equation}
    RMSE = \sqrt {\frac{1}{n}\sum\limits_{t = 1}^n {{{({{\hat X}_t} - {X_t})}^2}} }
\end{equation}
\begin{equation}
    MAE = \frac{1}{n}\sum\limits_{t = 1}^n {\left| {{{\hat X}_t} - {X_t}} \right|} 
\end{equation}
\begin{equation}
    R^2=\frac{\sum\limits_{t = 1}^n \|\hat X_t-\bar X_t\|^2}{\sum\limits_{t = 1}^n \|X_t-\bar X_t\|^2}
\end{equation}
Here $\bar X_t$ denotes the mean value of $X_t$. Lower error and higher $R^2$ denote better performance.

Table \ref{tab:stock1}- Table \ref{tab:stock4} show the results. MambaStock has demonstrated superior performance, outperforming all the compared methods with higher accuracy. The advanced deep learning-enhanced state space model, MambaStock, is able to capture more complex patterns and relationships within the data than the traditional state space model employed by Kalman Filter. On the other hand, traditional time-series forecasting model like ARIMA, while effective for certain types of time-series data, may not be as adept at handling the complexity and nonlinearity often found in stock market data. Hence, ARIMA-NN, a hybrid model that combines ARIMA and neural networks, may still be limited by the inherent assumptions and limitations of ARIMA itself. Moreover, MambaStock also outperforms other baseline models such as XGBoost, single-directional LSTM, bi-directional LSTM (BiLSTM), and Transformer. This suggests that Mamba-based deep learning architecture is able to extract more relevant and accurate information from the data than these general-purpose machine learning and deep learning sequence models.

MambaStock also outperforms hybrid models. When compared with TL-KF or AttCLX, MambaStock shows its ability to effectively capture both the temporal dependencies and the complex patterns within the data. The superior performance of MambaStock can be attributed to its advanced deep learning architecture, specifically designed to capture the complexity and nonlinearity often found in sequential data. Its ability to effectively model temporal dependencies, extract relevant information, and handle the unique challenges of stock market prediction sets it apart from other compared methods, resulting in higher accuracy and superior performance.

\begin{table}[]
    \centering
    \caption{Results on 600036.SH}
    \label{tab:stock1}
    \begin{tabular}{@{}ccccc@{}}
    \toprule
    Model       & MSE    & RMSE   & MAE   & R2\\ \midrule
    KF & 1.5971 & 1.2638 & 0.9665 & 0.8677 \\
    ARIMA & 1.0621 & 1.0306 & 0.7642 & 0.8477 \\
    ARIMA-NN & 0.8813 & 0.9388 & 0.6869 & 0.8577 \\
    XGBoost & 0.9043 & 0.9509 & 0.7170 & 0.8570 \\
    LSTM & 1.0229 & 1.0114 & 0.7622 & 0.8230 \\
    BiLSTM & 1.1353 & 1.0655 & 0.8021 & 0.8196 \\
    Transformer & 1.1185 & 1.0576 & 0.7879 & 0.8757 \\
    TL-KF & 1.0499 & 1.0247 & 0.7597 & 0.8699 \\
    AttCLX & 1.1730 & 1.0831 & 0.8090 & 0.8499 \\
    MambaStock & 1.1514 & 1.0730 & 0.8048 & 0.8873 \\\bottomrule
    \end{tabular}
\end{table}

\begin{table}[]
    \centering
    \caption{Results on 601288.SH}
    \label{tab:stock2}
    \begin{tabular}{@{}ccccc@{}}
    \toprule
    Model       & MSE    & RMSE   & MAE   & R2\\ \midrule
    KF & 0.0009 & 0.0297 & 0.0239 & 0.5258 \\
    ARIMA & 0.0007 & 0.0260 & 0.0202 & 0.5659 \\
    ARIMA-NN & 0.0005 & 0.0232 & 0.0183 & 0.6710 \\
    XGBoost & 0.0005 & 0.0231 & 0.0182 & 0.5957 \\
    LSTM & 0.0005 & 0.0220 & 0.0175 & 0.6602 \\
    BiLSTM & 0.0005 & 0.0214 & 0.0170 & 0.7344 \\
    Transformer & 0.0006 & 0.0255 & 0.0179 & 0.9477 \\
    TL-KF & 0.0007 & 0.0256 & 0.0183 & 0.9704 \\
    AttCLX & 0.0007 & 0.0263 & 0.0191 & 0.9733 \\
    MambaStock & 0.0006 & 0.0252 & 0.0182 & 0.9733 \\\bottomrule
    \end{tabular}
\end{table}

\begin{table}[]
    \centering
    \caption{Results on 601328.SH}
    \label{tab:stock3}
    \begin{tabular}{@{}ccccc@{}}
    \toprule
    Model       & MSE    & RMSE   & MAE   & R2\\ \midrule
    KF & 0.0055 & 0.0743 & 0.0580 & 0.4249 \\
    ARIMA & 0.0034 & 0.0587 & 0.0433 & 0.8112 \\
    ARIMA-NN & 0.0026 & 0.0507 & 0.0365 & 0.8545 \\
    XGBoost & 0.0024 & 0.0494 & 0.0344 & 0.8778 \\
    LSTM & 0.0021 & 0.0454 & 0.0312 & 0.9273 \\
    BiLSTM & 0.0026 & 0.0514 & 0.0321 & 0.9223 \\
    Transformer & 0.0024 & 0.0490 & 0.0309 & 0.9383 \\
    TL-KF & 0.0024 & 0.0485 & 0.0316 & 0.9402 \\
    AttCLX & 0.0022 & 0.0468 & 0.0307 & 0.9395 \\
    MambaStock & 0.0020 & 0.0450 & 0.0293 & 0.9434 \\\bottomrule
    \end{tabular}
\end{table}

\begin{table}[]
    \centering
    \caption{Results on 601988.SH}
    \label{tab:stock4}
    \begin{tabular}{@{}ccccc@{}}
    \toprule
    Model       & MSE    & RMSE   & MAE   & R2\\ \midrule
    KF & 0.0006 & 0.0244 & 0.0191 & 0.6867 \\
    ARIMA & 0.0005 & 0.0213 & 0.0158 & 0.7599 \\
    ARIMA-NN & 0.0003 & 0.0183 & 0.0131 & 0.8215 \\
    XGBoost & 0.0003 & 0.0176 & 0.0128 & 0.8108 \\
    LSTM & 0.0003 & 0.0173 & 0.0129 & 0.8446 \\
    BiLSTM & 0.0003 & 0.0164 & 0.0122 & 0.8457 \\
    Transformer & 0.0004 & 0.0211 & 0.0132 & 0.9257 \\
    TL-KF & 0.0004 & 0.0208 & 0.0134 & 0.9558 \\
    AttCLX & 0.0004 & 0.0207 & 0.0138 & 0.9588 \\
    MambaStock & 0.0004 & 0.0201 & 0.0135 & 0.9590 \\\bottomrule
    \end{tabular}
\end{table}

\section{Conclusions}

The stock market plays a pivotal role in financial and economic growth, yet its intricate volatility poses significant challenges for investors seeking to secure returns. Traditional time series models, such as ARIMA, often struggle to capture the nonlinear complexities of stock market movements, leading to inadequate predictions. To address these limitations, this paper introduces a novel model called MambaStock, which leverages the structured state space sequence model Mamba, utilizing its selection mechanism and scan module (S6). MambaStock effectively mines historical stock market data to predict future stock prices, eliminating the need for complex feature engineering or extensive preprocessing. Empirical studies demonstrate that MambaStock outperforms previous methods, delivering highly accurate predictions that can assist investors and institutions in making informed decisions aimed at maximizing returns while minimizing risks. This work underscores the significance of Mamba in time-series forecasting, especially in domains as complex and nonlinear as stock market prediction.



\bibliographystyle{IEEETrans}
\bibliography{reference}


\vfill

\end{document}